\documentclass[a4paper,10pt]{article}
\usepackage{graphicx}
\usepackage{amsmath}
\usepackage{amsfonts}
\usepackage{amssymb}
\usepackage{epsfig}
\newcommand{\be}{\begin{equation}}
\newcommand{\ee}{\end{equation}}

\newcommand{\no}{\nonumber\\}
\newcommand{\ba}{\begin{eqnarray}}
\newcommand{\ea}{\end{eqnarray}}
\newcommand{\bg}{\begin{multline}}
\newcommand{\eg}{\end{multline}}

\newcommand{\eint}{\int_{\sigma_1}^{\infty} dE\, }
\def\gl#1{(\ref{#1})}
\def\tr#1{\mbox{\rm tr}\left\{#1\right\}}

\begin{document}

\title{Spontaneous $P$-violation in QCD in extreme conditions}
\bigskip

\small{\author{ \small
A.A.~Andrianov\footnote{andrianov@ecm.ub.es}, D.~Espriu\footnote{espriu@ecm.ub.es},\\
\small Departament d'Estructura i Constituents de la Mat\`eria and\\
\small  ICCUB Institut de Ci\`encies del Cosmos,\\
\small Universitat de Barcelona, Diagonal 647, 08028 Barcelona, Spain\\
\small V.A.~Andrianov\footnote{v.andriano@rambler.ru}\\
\small V.A. Fock Department of Theoretical Physics, St. Petersburg State University,\\
\small  ul. Ulianovskaya 1, 198504, St.Petersburg, Russia}}
\date{}
\maketitle

\begin{abstract}
We investigate the possibility of parity being spontaneously violated in QCD at finite baryon density
and temperature. The analysis is done for an idealized homogeneous and infinite nuclear matter
where the influence of density can be examined with the help of constant chemical potential.
QCD is approximated by a generalized $\sigma$ model with two isomultiplets of scalars and
pseudoscalars. The interaction with the chemical potential is introduced via the coupling
to constituent quark fields as nucleons are not considered as point-like degrees of freedom
in our approach. This mechanism of parity violation is based on interplay between lightest
and heavier degrees of freedom and it cannot be understood in simple models retaining the
pion and nucleon sectors solely. We argue that, in the appropriate environment
(dense and hot nuclear matter of a few normal densities and moderate temperatures), parity
violation may be the rule rather than the exception and its occurrence is well compatible
with the existence of stable bound state of normal nuclear matter. We prove that the so
called 'chiral collapse' never takes place for the parameter region supporting
spontaneous parity violation. \end{abstract}

%\pacs{11.30.Qc,11.30.Er,12.38.Aw,12.40.Yx}

%\maketitle
\bigskip

\noindent {\hfill\normalsize\tt Preprint\quad UB-ECM-PF/09-09;\quad ICCUB-09-188}
\bigskip

\section{\label{intro} Introduction}

The appearance of parity ($P$) violation via pseudoscalar condensation for sufficiently
large values of temperature and/or chemical potential has been attracting much interest
during last decades to search it both in dense nuclear matter (in neutron/quark stars
and heavy ion collisions at intermediate energies) and in strongly interacting
quark-gluon matter (``quark-gluon plasma'' in heavy ion collisions at very high energies).
At finite baryon density it was conjectured by A. Migdal in \cite{migdal} long ago
(and revisited in \cite{picon}). One should also mention the possibility of $(C)P$-parity
violation in meta-stable nuclear bubbles created in hot nuclear matter \cite{kharzeev}.
Finally $P$ violation might conceivably accompany the transitions to open color phases\cite{scon}
such as CFL (color-flavor locking) or SC (superconducting), but these are phases beyond
the range of validity of our analysis. While it was argued in \cite{witten} that parity,
and vector flavor symmetry could not undergo spontaneous symmetry breaking in a vector
like theory such as QCD, the conditions under which the results of \cite{witten} hold
(positivity of the measure) are not valid for non-zero chemical potential.

Parity violation in QCD would lead to rather remarkable experimental signals such as
the same in-medium resonance being able to decay into even and odd number of pions,
the presence of additional Goldstone bosons (six right at the phase transition
in the exact chiral limit, and five throughout the broken parity phase), changes
in the nuclear equation of state, and isospin breaking effects in the pion decay constant
and substantial modification of the weak decay constant $F_{\pi'}$ for massless charged
pions, giving an enhancement of electroweak decays.

In this work we shall explore further the interesting possibility of spontaneous parity
violation employing effective Lagrangian techniques in the range of nuclear densities
where the hadron phase persists and quark percolation does not occur yet. Our effective
Lagrangian is a realization of the generalized linear $\sigma$ model, but including the
two lowest lying resonances in each channel, those that are expected to play a role
in this issue. As it will hopefully become clear later this is the minimal model
where this interesting possibility can be realized. The use of effective Lagrangians
is also crucial to answer the second question of interest, namely how would parity violation
originating from a finite baryon density eventually reflect in hadronic physics.

This work is a continuation of the one published by two of the present authors in \cite{ae}.
Here we extend the analysis previously done to the case of non-zero temperature
showing that the parity violation phase persists in some finite domain in the $\mu-T$ plane.
How far up in $T$ this domain extends is not clear yet, as eventually one is expected
to enter in a deconfined phase for which the effective meson-quark Lagrangian is not valid.

We have considered some of the thermodynamic properties of dense baryon matter in our approach
and we have also addressed the issue of how our model can describe the saturation point
and the formation of stable nuclear matter. We had not touched upon this issue in detail
in \cite{ae}. We find that our description turns out to be rather accurate in describing
nuclear matter formation avoiding the unacceptable 'chiral collapse'\cite{buballa1}
usually present in quark  models.

We have also examined departures from the chiral limit, i.e. allowing for non-zero quark
masses. This leads to rather interesting results as in this case the usual pions are
not exactly massless, but the new Goldstone bosons appearing at the transition point
to the parity violating phase are. Strong interaction phenomenology becomes indeed very unfamiliar
at that point. These results along with other considerations shall be summarized
in a separate publication \cite{aae}.

Many techniques have been used to study QCD in unusual conditions: from
meson-nucleon\cite{migdal,picon,bary,takah} or quark-meson \cite{sigqu,grei} lagrangians
for low-dense nuclear matter to models of Nambu-Jona-Lasinio type \cite{NJL,ebert}
for high-dense quark matter \cite{scon}. However, for reasons explained below, all
hadronic models lack for one reason or another some essential ingredient.
One should also mention the extensive lattice work, plagued with  technical
difficulties when $\mu\neq 0$ \cite{philip}. Let us finally comment
that the range of intermediate nuclear densities (from 3 to 10 times the usual nuclear density)
where we expect parity breaking to occur is of high interest as it may be reached both
in compact stars \cite{grei} and heavy-ion collisions \cite{heavyion}.

\section{\label{sigma-model} A generalized sigma model for QCD}

The simplest hadronic effective theory is the  linear $\sigma$-model of Gell-Mann and
Levy\cite{gellmann}, which contains a multiplet of the lightest isoscalar $\sigma$ and
isotriplet pseudoscalar $\pi^a$ fields. Spontaneous chiral symmetry breaking emerges due
to a non-zero value for $\langle \sigma\rangle  \sim \langle \bar q q \rangle /
 F^2_\pi$. Current algebra techniques indicate that in order to relate
this model to QCD one has to choose a real condensate for the scalar density,
with its sign opposite to current quark masses, and avoid any parity violation due to a
v.e.v. of the pseudoscalar density. The introduction of a chemical potential does not change
the phase of the condensate and therefore does not generate any spontaneous parity violation.

Thus too simple phenomenological models retaining only the lightest degrees of freedom are
not capable to explore all the different phases that the presence of manifest $CP$
violation due to the non-zero chemical potential opens. The need to include more resonances
can be qualitatively motivated by the fact that at substantially larger densities typical
distances between baryons are shrinking considerably and meson excitations with Compton
wave-lengths much shorter than the pion one start playing an important role.

The minimal model having the possibility of describing spontaneous parity breaking (SPB)
contains two multiplets of scalar/pseudoscalar fields
$H_j = \tilde{\sigma}_j {\bf I} + i \hat\pi_j, \quad j = 1,2$, with
$\hat\pi_j \equiv \tilde{\pi}^a_j \tau^a$ where $\tau^a$ are Pauli matrices.
We require an exact $SU(2)_L \times SU(2)_R$ symmetry in the chiral limit.
We should think of these two chiral multiplets as representing the two lowest-lying
radial states for a given $J^{PC}$. Adding more resonances would surely provide a
better description of the hadronic phase of QCD, but the present model already
possesses all the necessary ingredients to study SPB. Inclusion of higher-mass states is
required at substantially larger densities. Likewise we do not need to include vector
resonances (with one exception to be discussed below).

The effective potential of this generalized $\sigma$ model
\ba
V_{\text{eff}}&=& \frac12 \tr{- \sum_{j,k=1}^2 H^\dagger_j \Delta_{jk} H_k +
\lambda_1 (H^\dagger_1 H_1)^2 + \lambda_2 (H^\dagger_2 H_2)^2+
\lambda_3 H^\dagger_1 H_1 H^\dagger_2 H_2 \right.\no && \left.+
\frac12 \lambda_4 (H^\dagger_1 H_2 H^\dagger_1 H_2 + H^\dagger_2 H_1 H^\dagger_2 H_1) +
\frac12 \lambda_5 (H^\dagger_1 H_2 + H^\dagger_2 H_1) H^\dagger_1 H_1  \right.\no &&\left.+
\frac12 \lambda_6 (H^\dagger_1 H_2 + H^\dagger_2 H_1) H^\dagger_2 H_2  }
+ {\cal O}(\frac{|H|^6}{\Lambda^2}) \label{effpot1}
\ea
contains 9 real constants. QCD bosonization rules imply that they are $\sim N_c$.
The neglected terms will be suppressed at least by inverse powers of the chiral symmetry
breaking (CSB) scale $\Lambda \simeq 1.2$ GeV. If we assume the v.e.v. of $H_j$ to be of
the order of the constituent mass $0.2 \div 0.3$ GeV,  it is reasonable to neglect these terms.

We take $ H_1$ as the chiral multiplet coupling locally to the quark fields
(see section \ref{finite-mu}). Using the global invariance of the model we can parameterize
\ba
H_1 (x) = \sigma_1 (x) \xi^2(x)= \sigma_1 (x) \exp\left(i\frac{\pi_1^a \tau_a}{F_0}\right) ;
\quad H_2 (x) = \xi (x)\Big(\sigma_2 (x) +i \hat\pi_2 (x)\Big)\xi (x) \label{chipar}.
\ea
The parities of $\sigma_2 (x)$ and $\hat\pi_2$ are even and odd, respectively (in the absence of SPB).
The potential \gl{effpot1} reads \ba V_{\text{eff}}&=& - \sum_{j,k=1}^2 \sigma_j \Delta_{jk} \sigma_k -
\Delta_{22} (\pi_2^a)^2 +  \lambda_2  \Big((\pi_2^a)^2\Big)^2   +
(\pi_2^a)^2 \Big((\lambda_3 - \lambda_4) \sigma_1^2 + \lambda_6 \sigma_1 \sigma_2 +
2\lambda_2 \sigma_2^2 \Big)\label{efpot1}\no && + \lambda_1 \sigma_1^4
 + \lambda_2 \sigma_2^4
+ (\lambda_3 + \lambda_4) \sigma_1^2 \sigma_2^2 +
\lambda_5 \sigma_1^3 \sigma_2 + \lambda_6 \sigma_1 \sigma_2^3 .
\ea

The corresponding gap equations are
\ba
2(\Delta_{11} \sigma_1 + \Delta_{12} \sigma_2) &=& 4 \lambda_1 \sigma_1^3 +
3\lambda_5 \sigma_1^2 \sigma_2 + 2 (\lambda_3 + \lambda_4) \sigma_1 \sigma_2^2  +
\lambda_6 \sigma_2^3 + \rho^2 \Big(2(\lambda_3 - \lambda_4) \sigma_1 + \lambda_6 \sigma_2\Big)
\label{mg1}\\
2(\Delta_{12} \sigma_1 +\Delta_{22} \sigma_2) &=& \lambda_5 \sigma_1^3 + 2 (\lambda_3 +
\lambda_4)  \sigma_1^2 \sigma_2 + 3 \lambda_6  \sigma_1 \sigma_2^2 + 4 \lambda_2 \sigma_2^3  +
\rho^2 \Big(\lambda_6  \sigma_1 +4 \lambda_2 \sigma_2 \Big),
\label{mg2}\\
0&=&\rho\Big( - \Delta_{22} +(\lambda_3 - \lambda_4) \sigma_1^2 +
\lambda_6 \sigma_1 \sigma_2+ 2\lambda_2 \sigma_2^2 + 2  \lambda_2 \rho^2\Big) , \label{mg3}
\ea
where the following notation has been introduced:
$\langle\pi_1^a\rangle=\langle\pi^0\rangle\delta^{0a}, \langle\pi_2^a\rangle=\rho\delta^{0a}$.

The above effective potential must exhibit the usual chiral symmetry breaking pattern at $\mu=T=0$.
For this to happen $\langle\sigma_1\rangle$ must acquire a real and positive v.e.v.
to agree with current algebra considerations. Note that $\langle \pi^0\rangle$
does not appear at all in the gap equations and hence its value is completely undetermined,
but from (\ref{chipar}) we see that a non zero value for $\langle \pi^0\rangle$
would affect the phase of the scalar condensate. The addition of a small mass for the
quarks fixes the phase of the breaking without having to appeal to other arguments \cite{aae}.

The previous set of gap equations may have several solutions for $\sigma_1$ and $\sigma_2$,
but since we know that in normal conditions QCD does not break parity,
$\rho$ must vanish. Since for the potential to be well defined requires $\lambda_2 >0 $, a
{\it sufficient} condition for the absence of SPB is
\be
(\lambda_3 - \lambda_4) \sigma_1^2 + \lambda_6 \sigma_1 \sigma_2 + 2\lambda_2 \sigma_2^2  > \Delta_{22}.
\label{ineq1}
\ee
On the other hand, the mass of the pseudoscalar $\pi_2$ is governed by the second variation
\be
V^{(2)}_{\pi_2\pi_2}= 2(-\Delta_{22}+(\lambda_3 - \lambda_4) \sigma_1^2 +
\lambda_6 \sigma_1\sigma_2 + 2\lambda_2 \sigma_2^2+ 6\lambda_2\rho^2). \label{secvar11}
\ee
Positivity of this mass for $\rho=0$ implies (\ref{ineq1}). The condition is
therefore {\it necessary} too.

Let us establish the necessary and sufficient conditions for CSB to take place in normal
conditions. The necessary condition to have a minimum of $V_{\text{eff}}$ for non-zero
$\sigma_j$ (for vanishing $\rho$) can be derived from the condition to get a local maximum
(or at least a saddle point) for zero $\sigma_j$. This extremum is characterized by the matrix
$-\Delta_{ij}$ in (\ref{efpot1}) which must have at least one negative eigenvalue.
The sufficient condition follows from the positivity of the second variation of $V_{\text{eff}}$
for a non-trivial solution of the two first equations \gl{mg1}, \gl{mg2} at $\rho = 0$.

The matrix containing the second variations in the scalar sector is $V^{(2)}$
\ba
\frac12 V^{(2)}_{\sigma_1\sigma_1} &=& -\Delta_{11} + 6\lambda_1 \sigma_1^2 +
3\lambda_5 \sigma_1 \sigma_2 + (\lambda_3 + \lambda_4)  \sigma_2^2 , \no
V^{(2)}_{\sigma_1\sigma_2} &=& - 2\Delta_{12} + 3 \lambda_5 \sigma_1^2 +
4  (\lambda_3 + \lambda_4) \sigma_1 \sigma_2 + 3\lambda_6 \sigma_2^2  , \no
\frac12 V^{(2)}_{\sigma_2\sigma_2} &=& -  \Delta_{22} +
(\lambda_3 + \lambda_4)  \sigma_1^2 + 3 \lambda_6  \sigma_1 \sigma_2 + 6 \lambda_2 \sigma_2^2 .
\label{secvar}
\ea
The required conditions are given by $\tr{ \hat V^{(2)}} >0$ and
 $\mbox{\rm Det} \hat V^{(2)} >0$, where $\hat V^{(2)}$ is the matrix of second
variations restricted to the scalar fields. For positive matrices this means
$ V^{(2)}_{\sigma_j\sigma_j} >0$. The eigenvalues of \gl{secvar} eventually give the masses
squared of the scalar mesons and thereby must be positive.

\section{\label{landscape}The landscape of solutions}

The above gap equations exhibit a rather complex landscape of extrema and it is important
to simplify the discussion as much as possible. Thus we shall consider here the case where
the eigenvalues of the matrix $-\Delta_{ij}$ in (\ref{efpot1}) are all negative.
This corresponds to the origin being a local maximum which appears as the most natural condition.
For the sake of searching for extrema the effective potential can be further simplified
by making a general linear transformation on the $H_j$ fields
\be
\tilde H_j= \sum_{k=1,2} L_{jk} H_k. \label{lintra}
\ee
The $L_{jk}$ must be real to preserve the reality of $\sigma_j$ and $\pi_j$.
This transformation has four real parameters which are enough to set $\Delta_{jk}=\Delta \delta_{jk}$
and $\lambda_5=0$ (for a more detailed discussion see \cite{aae}). Thus the particular
choice of $\Delta_{12} = \lambda_{5}=0$ does not make the analysis of the extrema less general.
Apart from $\lambda_5$, which disappears, the other effective coupling get modified
by the above linear transformations: $\lambda_i\to\tilde\lambda_i$ .
To avoid complicating the notation we shall omit the tilde for the transformed
fields and constants in the subsequent formulae.

One of the solution of the simplified equations is $\sigma_2 = 0$ and
$\sigma_1^2 = \Delta/2 \lambda_1$. This is a minimum, as it follows from
\eqref{secvar11} and \eqref{secvar},
provided that $\lambda_3 \pm \lambda_4 > 2\lambda_1> 0$. Note that the last inequality
follows from the stability of the potential.  When $\sigma_2 = 0$ the effective potential
exhibits an enlarged symmetry $Z_2 \times Z_2$ as $\lambda_6$ does not contribute.

If $\sigma_2 \not= 0$, by combining the first two gap equations it is possible to determine
another set of solutions. They are given in term of the ratio
$x\equiv \sigma_2/\sigma_1$ by the solutions to the cubic equation
\be
2\lambda_1 - (\lambda_3+\lambda_4) - \frac32 \lambda_6 x +
(\lambda_3+\lambda_4-2\lambda_2)x^2 + \frac12 \lambda_6 x^3=0. \label{mgratio}
\ee
This equation may have one or three real roots. Taking into account that $\lambda_6$ must be
taken negative (see below) and that the zero-order term is negative and the term linear in $x$
is positive we conclude that there are  either two positive and one negative real root
or just one negative real root.

In order to determine the nature of these extrema we need to compute the matrix of second
variations. After solving for $\Delta$ using the gap equations and taking into account that
$\sigma_2\neq 0$ the condition for a local minimum reads
 \ba
\frac12 V^{(2)}_{\sigma_1\sigma_1} &=& 4\lambda_1 (\sigma_1)^2 - \frac12 \lambda_6 \frac{\sigma_2^3}{\sigma_1}
> 0, \no
V^{(2)}_{\sigma_1\sigma_2} &=&  4  (\lambda_3 + \lambda_4) \sigma_1 \sigma_2 + 3\lambda_6 \sigma_2^2  , \no
\frac12 V^{(2)}_{\sigma_2\sigma_2} &=& \frac32 \lambda_6  \sigma_1 \sigma_2 +
4 \lambda_2 \sigma_2^2 > 0,\no
\frac12 V^{(2)}_{\pi_2\pi_2}&=&  - 2\lambda_4 \sigma_1^2 - \frac12 \lambda_6 \sigma_1\sigma_2  > 0.
\label{secvar1}
\ea
The first element is positive provided that $\lambda_6 < 0$, the last one is also positive if
$4\lambda_4 \sigma_1 < -\lambda_6 \sigma_2$ which follows
directly from the gap equations themselves and the condition for the absence of SPB at
$\mu=0$ (\ref{ineq1}). As for the third one, this gives another condition
\be
\lambda_2 \sigma_2 > -\frac38 \lambda_6  \sigma_1\qquad \Rightarrow\, x >0.
\ee
Therefore, either there is only one saddle point (negative solution) or one minima and two saddle points
(one negative, two positive solutions).

As the effective potential is symmetric against the transformation $\sigma_j \rightarrow - \sigma_j$,
there is actually a doubling of the solutions. Taking into account the solutions with $\sigma_2=0$ this
makes a maximum of eight extrema; actually nine if we include the maximum at $\sigma_j=0$.
Comparison with chiral algebra arguments fix the physical sign of $\sigma_j$ thus breaking the
degeneracy. Of these extrema only two of them at most are minima. As to which of the two minima
(should they exist simultaneously) has lower energy, this is actually
depending on the specific values of the constants. From now on we revert to the original set of fields
and coupling constants undoing the linear transformation (\ref{lintra}).

\section{\label{finite-mu} Inclusion of temperature and baryon chemical potential}

After bosonization the baryon chemical potential $\mu$ is transmitted to the meson sector
(in the leading order of chiral expansion) via a local quark-meson coupling.
In turn, in the large $N_c$ limit one can neglect the temperature dependence due to meson
collisions and assume that the temperature $T$ is induced with the help of the imaginary time
Matsubara formalism for Green functions - Matsubara frequencies for quarks
$ \omega_n = (2n+1)\pi/\beta$ with  $\beta = 1/kT$. In the real world with 3 colors
this is of course an approximation, but nevertheless it should be sufficient to describe
qualitatively the interplay between baryon density and temperature, and it is the one consistent
with our mean field approach anyway.

As already mentioned we take the chiral multiplet to have local couplings with
the quark fields as being $H_1$. The set of coupling constants in \eqref{effpot1}
allows us to fix the Yukawa coupling constant to unity. Thus $\mu$ and $T$ are transmitted
to the boson sector by the term
\be
\Delta {\cal L}= - (\bar q_R H_1 q_L +\bar q_L H_1^\dagger q_R)\longrightarrow - \bar q \sigma_1 q,
\label{yukawa}
\ee
where $q_{L,R}$ are assumed to be constituent quarks. We do not include baryon fields explicitly
and therefore quark matter and nuclear matter are indistinguishable in our approach.
Of course in the conditions where unconfined quark matter would be present the density
would be such that our model is not applicable anymore.

After integrating out the constituent quarks the full temperature and chemical potential dependence,
to the leading orders in chiral expansion, find their way into the gap equations. Namely \gl{mg1}
is modified to
\be
2(\Delta_{11} \sigma_1 + \Delta_{12} \sigma_2) = 4 \lambda_1 \sigma_1^3 + 3\lambda_5 \sigma_1^2 \sigma_2 +
2 (\lambda_3 + \lambda_4) \sigma_1 \sigma_2^2  + \lambda_6  \sigma_2^3 +
\rho^2 \Big(2(\lambda_3 - \lambda_4) \sigma_1 + \lambda_6  \sigma_2\Big) +
2{\cal N}\sigma_1 {\cal A}(\sigma_1, \mu, \beta),\label{var1}
\ee
with $ {\cal N} \equiv \frac{N_{c}N_f}{4\pi^2}$,  and where
\ba
{\cal A}(\sigma_1, \mu, \beta) &=&\frac{1}{\beta}
\sum_n \exp(i\omega_n\eta) \int \frac{d^3p}{\pi} \frac{1}{(i\omega_n+\mu)^2-p^2 -\sigma_1^2} - [T=0,\ \mu = 0]\no
&=& 2 \eint \, \sqrt{E^2-\sigma_1^2} \, \frac{\cosh(\beta\mu) +
\exp(-\beta E)}{\cosh(\beta\mu) + \cosh(\beta E)} \label{Aend} \ ,
\ea
where the Fermi distribution has been introduced. ${\cal A}$ originates from the one-loop
 contribution to $V_{\text{eff}}$
\be
\Delta V_{\text{eff}}(\sigma_1, \mu, \beta) =
- \frac43 {\cal N}\int\limits_{\sigma_1}^{\infty}\, dE \Bigl(E^2-\sigma_1^2\Bigr)^{3/2} \,
\frac{\cosh(\beta\mu) + \exp(-\beta E)}{\cosh(\beta\mu) + \cosh(\beta E)},
\ee
normalized to vanish for very large $\sigma_1$. All the dependence on the environment
is in the function ${\cal A}(\sigma_1,\mu,\beta)$.

Let us particularize to the zero-temperature case. At $T = 0$
\ba
\Delta V_{\text{eff}}(\mu) &=& \frac{{\cal N}}{2}
\Theta(\mu-\sigma_1)\Biggl[\mu\sigma_1^2\sqrt{\mu^2-\sigma_1^2}-
\frac{2\mu}{3}(\mu^2-\sigma_1^2)^{3/2}- \sigma_1^4\ln{\frac{\mu+\sqrt{\mu^2-\sigma_1^2}}{\sigma_1}}\Biggr],
\label{potmu}
\ea
and
\be
{\cal A}(\sigma_1, \mu, \beta=\infty ) = 2 \theta(\mu - \sigma_1) \int^\mu_{\sigma_1} dE \,
\sqrt{E^2-\sigma_1^2} = \mu\sqrt{\mu^2-\sigma_1^2}- \sigma_1^2\ln{\frac{\mu+\sqrt{\mu^2-\sigma_1^2}}{\sigma_1}}
\ee
Using the gap equations \eqref{mg2}, \eqref{mg3} and \eqref{var1}, the value of the effective potential
at its minima is given by the compact expression
\ba
V_{\text{eff}}(\mu) = - \frac12 \sum_{j,k=1}^2 \sigma_j(\mu) \Delta_{jk} \sigma_k(\mu) -
\frac12 \Delta_{22} \rho^2(\mu) -
\frac{{\cal N}}{3} \mu \Big(\mu^2 - \sigma_1(\mu)^2\Big)^{3/2} \theta\Big(\mu - \sigma_1(\mu)\Big).
\label{effmu}
\ea
It should be emphasized that all the above results have corrections of
${\cal O}\left(\mu^2/\Lambda^{2},\sigma_1^2/\Lambda^{2}\right)$.

\section{Thermodynamic properties of the model}

Thermodynamically the system is described by the pressure $p$ and the energy density,
$\varepsilon$. The pressure is determined by the potential density difference with and without
the presence of chemical potential, $dp = -dV$,
\be
p(\sigma_j(\mu), \mu) \equiv V_{\text{eff}}\Big(\sigma_j^0\Big)-
V_{\text{eff}}\Big(\sigma_j(\mu),\rho(\mu), \mu\Big), \label{press}
\ee
where the dependence of $\sigma_j$ and  $\rho(\mu)$ on $\mu$ has been shown explicitly and
$\sigma_j^0\equiv \sigma_j(0)$. The energy density is related to the pressure by
\be
\varepsilon = - p +N_c \mu \varrho_B . \label{enden}
\ee
The chemical potential is defined as
\be
\partial_{\varrho_B} \varepsilon = N_c \mu , \label{chem}
\ee
with the entropy and volume held fixed. The factor $N_c$ is introduced to relate
the quark and baryon chemical potentials. Since $\varepsilon$ is independent of $\mu$,
\be
\partial_\mu p = N_c \varrho_B. \label{varrho}
\ee
Thus the relation between baryon density, Fermi momenta and the chemical potential is for quark matter
\ba
\varrho_B = -\frac{1}{N_c} \partial_\mu  V_{\text{eff}} =
\frac{N_f}{3\pi^2} p_F^3 = \frac{N_f}{3\pi^2} (\mu^2- \sigma_1(\mu)^2)^{3/2}.
\label{rhomu}
\ea
This set of identities provides a functional relation between $\varrho_B$ and $\mu$.
The pressure can also be written as
\be
p = \varrho_B^2 \partial_{\varrho_B}\left(\frac{\varepsilon}{\varrho_B}\right),
\ee
showing that the energy per baryon has an extremum when the pressure vanishes.
Since the pressure is an increasing function of the density as we have seen
(and it obviously vanishes at zero density), and infinite nuclear matter is stable
(thus implying zero pressure too) the phase diagram in the $p,\varrho_B$ plane
must necessarily exhibit a discontinuity. This would correspond to the first
order transition associated to the formation of nuclear matter at some critical value $\mu^*$.
This implies that several solutions of the gap equations must coexist around $\mu^*$.

Our model consisting of two scalar isomultiplets is somewhat too
simple in one respect. The stabilization of nuclear matter requires not only attractive scalar
forces (scalars) but also repulsive ones (vector-mediated) \cite{wal}. Conventionally,
the latter ones are associated to the interactions mediated by the iso-singlet vector
$\omega$ meson. Let us supplement our action with the free $\omega$ meson lagrangian
and its coupling to quarks
\be
\Delta {\cal L}_\omega = - \frac14 \omega_{\mu\nu}\omega^{\mu\nu} +
\frac12 m_\omega^2 \omega_\mu \omega^\mu -
g_{\omega\bar q q} \bar q \gamma_\mu \omega^\mu q, \label{omega}
\ee
with a coupling constant $g_{\omega\bar q q}\sim {\cal O} (1/\sqrt{N_c})$. After bosonization of QCD,
on symmetry grounds, any vector field interacts with scalars in the form of commutator
and therefore $\omega_\mu $  does not show up in the effective potential $H_j$ fields at the
lowest order. However in the quark sector the time component $\omega_0$ interplays with
the chemical potential and it is of importance to describe the dense nuclear matter properties.
Let us assign a constant v.e.v. for this component
$g_{\omega\bar q q} \langle\omega_0 \rangle \equiv \bar\omega$. Then one needs to compute
the modification of the effective potential due to the replacement
$\mu \rightarrow \mu + \bar\omega\equiv \bar\mu$. The variable $\bar\omega $,
and accordingly $\bar\mu$, is dynamical and it also appears quadratically in the mass
term in \eqref{omega} which reads
\be
\Delta V_\omega = - \frac12 m_\omega^2\langle \omega_0^2\rangle =
-\frac12 \frac{(\bar\mu - \mu)^2}{G_\omega}, \quad G_\omega \equiv
\frac{g_{\omega\bar q q}^2}{m_\omega^2}\simeq {\cal O} (\frac{1}{N_c}) .\label{effomega}
\ee
The term \eqref{effomega} supplements the effective potential \eqref{effmu}:
$\bar V_{\text{eff}}(\mu) \equiv
 V_{\text{eff}}(\bar \mu) + \Delta V_\omega (\bar\mu, \mu)$ . $\bar V_{\text{eff}}(\mu)$ should
henceforth be used
in all the previous thermodynamical formulae. The replacement $\mu\to \bar\mu$ makes all
expectation values depend rather on $\bar\mu$, although of course the 'physical'
$\mu$ is the one directly related to the density via (\ref{rhomu}). $\bar\mu$
can be determined via the variation of $\bar  V_{\text{eff}}$ and is given by
\be
\frac{\bar\mu - \mu}{G_\omega} = - N_c \varrho_B (\mu) =
-\frac{N_c N_f}{3\pi^2}(\bar\mu^2- \sigma_1(\bar\mu)^2)^{3/2}. \label{physmu}
\ee

%The (in)compressibility in the quark matter is defined as $ K(\mu) =
% \partial_{\varrho_B} p $, where the derivative is made with the help of the function
%$\varrho_B(\mu)$ defined in (\ref{physmu}). For a zero pressure state (such as stable
%nuclear matter) this equals
%\be
%K= \varrho_B^2 \partial^2_{\varrho_B}\left(\frac{\varepsilon}{\varrho_B}\right)\bigl|_{p=0},
%\ee
%which must be positive since it corresponds to a minimum of the energy per baryon.
%In our model this is indeed the case
%\be
%K(\mu)=  N_c\varrho_B \partial_{\varrho_B} \mu=
%\frac{ p_F^2 (\bar\mu)}{\Big(\bar\mu-\sigma_1\partial_{\bar\mu} \sigma_1 (\bar\mu)\Big)}
%+  9 G_\omega \varrho_B (\mu)
%, \quad\partial_{\bar\mu}\sigma_1 = - 4 {\cal N}
%\sigma_1\sqrt{\bar\mu^2-\sigma_1^2} \, \frac{V_{\sigma_2\sigma_2}^{(2)}}{\det \hat V^{(2)}} < 0,
%\label{compress}
%\ee
%where $N_c = 3$ is taken.

\section{The saturation point and absence of chiral collapse}

A viable model of dense baryon matter must describe the phase transition to a stable bound
state at the usual density of infinite nuclear matter
$\varrho_B = \varrho_0 = 0.15\div 0.16 $ fm$^{-3}$, the so called ``saturation point''.
This phase transition is believed to be of first order similar to the vapor (hadron phase)
becoming saturated and turning into liquid (nuclear matter), when droplets of dense baryon
matter aggregate to form a homogeneous nuclear liquid.

However in simple quark models of the Nambu-Jona-Lasinio type \cite{buballa1} this
phase transition (for vanishing current quark masses) leads to a chirally symmetric phase
with zero dynamical mass (zero v.e.v. of scalar fields): i.e. the so called ``chiral collapse''
takes place. This is unphysical as chiral symmetry is not restored (at least not fully)
in nuclei. Furthermore in NJL-type models the typical baryon density
is substantially larger that the normal one (typically $\varrho_{B} \simeq 2.8 \varrho_0$).
For this reason the NJL model cannot be a reliable guide to phase transitions in dense nuclear matter.
Let us investigate whether our proposal avoids these pitfalls.

The saturation point where nuclear matter forms is characterized by vanishing pressure
as is the case for the vacuum. Given that the pressure is a non-decreasing function of $\mu$,
this can only happen if two solutions coexist and one of them, initially corresponding to
negative pressure, takes over. This is possible in our model because we can have up to two minima
both of them corresponding to chirally broken vacua. Let us see how this can be implemented.

If the value of the {\it effective} chemical potential corresponding to infinite nuclear
matter $\bar\mu^*$ is such that $\bar\mu^* <\sigma_1^0$ the one-loop correction
from the quark loop is zero for $\bar\mu < \bar\mu^*$ and $V_{\text{eff}}$ is constant
up to that point. Consequently, $p=0$ throughout this phase. Let us now assume that there
is another minimum that for $\mu=0$ has higher energy that the previous one, hence negative pressure,
and that this minimum is characterized by a value of $\sigma_1< \sigma_1^0$.
Then chemical potential corrections will start modifying this second solution as soon as
$\bar\mu> \sigma_1$. The pressure can only increase with $\mu$ and thus it will therefore
cross the $p=0$ line for a value $\bar\mu=\bar\mu^*$. At this point the second solution
takes over and $\bar\mu^*$ corresponds to the saturation point.
Let us denote by $\sigma_1^*$ the value of $\sigma_1$ at $\bar\mu^*$ obtained from this second solution.

Let us now prove that for the large set of coupling constants eventually leading to SPB (see below)
one of the fields at least has a non-zero expectation value, $\sigma_j \not=0$ in the chiral limit,
and chiral collapse is impossible. Indeed, suppose that $\sigma_{j} =0, \forall j$ at  $\mu^*$.
Then the matrix of second variations for the effective potential given in \eqref{secvar}
and taken from \eqref{potmu} reads
\ba
\frac12 V^{(2)}_{\sigma_ 1 \sigma_1} = -\Delta_{11} + {\cal N} \mu^2 , \
 V^{(2)}_{\sigma_1 \sigma_2} = - 2\Delta_{12}
 ,\ \frac12
V^{(2)}_{\sigma_2 \sigma_2} = -  \Delta_{22} . \label{secvar0}
\ea
In order to induce spontaneous breaking of parity one has to choose $\Delta_{22} > 0$ (see below).
Then from \eqref{secvar0} one finds that for any signs of other constants $\Delta_{11}, \Delta_{12}$
and for any value of $\mu$ the second variation matrix is never positive definite and one
finds a saddle point or a maximum at the presumed saturation point and beyond deep in the
nuclear matter phase. On the contrary, for $\Delta_{22} < 0$ one can always get a
large enough $\mu$ so that $\det\left[V^{(2)}\right]> 0,\ \tr{V^{(2)}} > 0 $ and chiral
collapse eventually takes place. As we have to guarantee the existence of stable nuclear matter
we assume $\Delta_{22} > 0$ from now on. Thus there seems to be a direct relation between
the absence of chiral collapse in the model and the eventual presence of a SPB phase.
Simplifying the model to have just one isomultiplet  (equivalent to NJL) simply corresponds
to considering the first element of the matrix of second variations and this changes sign
(implying chiral symmetry restoration) for $\mu$ large enough.

The energy crossing condition $p = 0$ at
$\bar\mu^\ast < \sigma_{1}^{0}, \sigma_j^\ast\equiv \sigma_{j}(\bar\mu^\ast)$ can be written,
taking into account \eqref{effmu} and \eqref{effomega}
\be
\sum_{j,k=1}^2 \Big(\sigma_j^0 \Delta_{jk} \sigma_k^0 - \sigma_j^\ast \Delta_{jk}\sigma_k^\ast\Big) =
\frac{N_cN_f}{6\pi^2} \bar\mu^\ast p_F^3 (\bar\mu^\ast) +
G_\omega \frac{N_c^2 N_f^2}{9\pi^4} p_F^6 (\bar\mu^\ast) =
\frac{N_c}{2} \bar\mu^\ast \varrho_B (\mu^\ast) + G_\omega N_c^2 \varrho_B^2 (\mu^\ast), \label{stablenm}
\ee
where $\bar\mu^\ast$ is related to the physical value of $\mu^\ast$ by Eq.\eqref{physmu}.
This relation represents the condition for the existence of symmetric nuclear matter.
It can always be fulfilled by an appropriate choice of $G_\omega$.

This is still a crude model and one certainly should not expect an extended sigma-quark model
to characterize normal nuclear matter with very good precision. Nevertheless taking this
at face value we can derive a relation between $\mu^*$ and $\sigma_1^*$:
$p_F = \sqrt{(\bar\mu^\ast)^2- (\sigma_1^\ast)^2} = 1.3\div 1.4\ \mbox{fm}^{-1} = 260\div 270$ MeV
that corresponds to $\varrho_0 = 0.15\div 0.16\ \mbox{fm}^{-3}$. Not bad.

\section{\label{SPB} The SPB phase transition}

We shall consider from now on the solution corresponding to the most stable minima for
$\bar\mu > \bar\mu^*$. To simplify the notation we shall also ignore the difference between
$\mu$ and the effective chemical potential $\bar\mu$.

The possibility of SPB is controlled by the inequality \gl{ineq1}. In order to approach a
SPB phase transition when the chemical potential is increasing we have to diminish the l.h.s.
of inequality \gl{ineq1} and therefore we need to have
\be
\partial_\mu\Big[(\lambda_3 - \lambda_4) \sigma_1^2
+ \lambda_6 \sigma_1 \sigma_2 + 2\lambda_2 \sigma_2^2\Big] < 0.
\ee
This is equivalent to
\ba
\Big(\lambda_6 \sigma_1 + 4\lambda_2\sigma_2  \Big) V^{(2)}_{\sigma_1 \sigma_2} <
\Big(2 (\lambda_3 - \lambda_4) \sigma_1+\lambda_6 \sigma_2\Big) V^{(2)}_{\sigma_2 \sigma_2}.
\label{ineq3}
\ea
This last inequality is a {\it necessary} condition that has to be satisfied by the model
it to be potentially capable of yielding SPB at large densities.

Let us examine the possible existence of a region of $\mu$ where $\rho\neq 0$. Then
\be
(\lambda_3 - \lambda_4) \sigma_1^2 +  \lambda_6 \sigma_1 \sigma_2 +2\lambda_2 \Big(\sigma_2^2 + \rho^2\Big)
= \Delta_{22} , \label{creq1}
\ee
After substituting $\Delta_{22}$ from \gl{creq1} into the second Eq.\gl{mg1} one finds that
\be
\lambda_5 \sigma_1^2 + 4 \lambda_4 \sigma_1 \sigma_2+ \lambda_6 \Big(\sigma_2^2
+ \rho^2\Big)  = 2\Delta_{12},\label{creq2}
\ee
where we have taken into account that $\sigma_1\neq 0$. Together with \gl{creq1}
this completely fixes the relation between v.e.v.'s of the scalar fields $ \sigma_{1,2}$
throughout the SPB phase independently of $\mu$ and $\rho$. If  $\lambda_2 \lambda_6 \not=0$
\gl{creq1} and \gl{creq2}  allow us to get rid of the v.e.v. $\rho$ and
\ba
\Big(2\lambda_5 \lambda_2 + \lambda_6(\lambda_4 - \lambda_3)\Big) \sigma_1^2 +
\Big(8 \lambda_2 \lambda_4 - \lambda_6^2\Big)\sigma_1 \sigma_2 =
4 \lambda_2 \Delta_{12} -  \lambda_6 \Delta_{22},
\ea
whose solution for $ \mu >\mu_{crit}$ is
\ba
\sigma_2 = A\sigma_1 + \frac{B}{\sigma_1},\qquad A \equiv \frac{2\lambda_5 \lambda_2 +
\lambda_6(\lambda_4 - \lambda_3)}{\lambda_6^2 - 8 \lambda_2 \lambda_4 }, \qquad
B \equiv \frac{\lambda_6 \Delta_{22}- 4 \lambda_2 \Delta_{12}}{\lambda_6^2 -8 \lambda_2 \lambda_4}.
\label{pbrrel}
\ea

Let us now  determine the critical value of the chemical potential,
namely the value $\mu_{crit}$
 where $\rho (\mu_{crit}) = 0$, but Eqs.\gl{creq1},
\gl{creq2}, \gl{pbrrel} hold. Combining the two equations \gl{creq1}, \gl{creq2}
\ba
(4\lambda_2 \Delta_{12} - \lambda_6 \Delta_{22}) x^2 + (2\lambda_6 \Delta_{12} -
4\lambda_4 \Delta_{22}) x + 2(\lambda_3 - \lambda_4)\Delta_{12}- \lambda_5 \Delta_{22} = 0,\quad
 x = \frac{\sigma_2}{\sigma_1}. \label{homeq}
\ea
In order for a SPB phase to exist this equation has to possess real solutions.
If $4\lambda_2 \Delta_{12} - \lambda_6 \Delta_{22}= 0$ there is
only one solution corresponding to a second order transition, but there may exist
other critical points  that fall beyond the accuracy of our low energy model
(which becomes inappropriate for small values of $\sigma_1$). We stress that
Eqs. \gl{pbrrel} and  \gl{homeq} contain only the constants of the potential and do
not depend on temperature and chemical potential manifestly.

Once we find $x_{crit} = x_{\pm}(\Delta_{12}, \Delta_{22}, \lambda_2,\ldots,\lambda_6)$
one can immediately calculate
\be
\sigma_1^\pm (\Delta_{jk}, \lambda_j) =\sqrt{\frac{B}{x_\pm - A}}, \qquad
\sigma_2^\pm (\Delta_{jk}, \lambda_j) = x_\pm \sigma_1^\pm . \label{sigmaone}
\ee
After substituting these values into Eq. \gl{var1} one derives the boundary of the $P$-violation phase
\be
{\cal N} {\cal A}(\sigma_1^\pm, \mu, \beta) = \Delta_{11}  -
2\lambda_1 (\sigma_1^\pm)^2 - \lambda_5 \sigma_1^\pm \sigma_2^\pm -
(\lambda_3 - \lambda_4) (\sigma_2^\pm)^2 , \label{strip}
\ee
which is a positive combination. The relation \gl{strip} defines a $P$-breaking divide line
in the $T - \mu$ plane. From \gl{Aend} one can obtain that ${\cal A} >0$ and
${\cal A} \rightarrow \infty$ when $T,\mu \rightarrow \infty$.
It means that for any nontrivial solution
$x_{\pm}, \sigma_{1}^\pm, \sigma_{2}^\pm$ ${\cal N}{\cal A} (\sigma_1^\pm, \mu, \beta) > 0$
the $P$-breaking phase boundary exists. If the phenomenon of $P$-violation
is realized for zero temperature it will take place in a domain involving lower
chemical potentials but higher temperatures.

\section{\label{spectrum} Nature of the SPB phase and physical spectrum}

Once a condensate for $\pi^{0}_2$ appears spontaneously the vector $SU(2)$ symmetry is
broken to $U(1)$ and two charged excited $\pi'$ mesons are expected to possess zero masses.
For simplicity let us consider zero temperature. After the second variations of the potential
we define the following quantities (one factor of $\rho$ is included in the definition of
${\cal V}$ for each derivation w.r.t. $\pi^{0}_2$ for convenience
\ba
\frac12 {\cal V}_{11} & \equiv & -\Delta_{11} + 6\lambda_1 \sigma_1^2  + 3\lambda_5 \sigma_1 \sigma_2 + (\lambda_3 + \lambda_4)  \sigma_2^2 + (\lambda_3 - \lambda_4) \rho^2 +  {\cal N}\left[\mu \sqrt{\mu^2-\sigma_1^2}- 3\sigma_1^2\ln{\frac{\mu+\sqrt{\mu^2-\sigma_1^2}}{\sigma_1}}\right], \no {\cal V}_{12}&\equiv& - 2\Delta_{12} + 3 \lambda_5 \sigma_1^2  + 4  (\lambda_3 + \lambda_4) \sigma_1 \sigma_2 + 3\lambda_6  \sigma_2^2 + \lambda_6 \rho^2, \no \frac12 {\cal V}_{22}&\equiv&-  \Delta_{22}  +  (\lambda_3 + \lambda_4)  \sigma_1^2 +
3 \lambda_6  \sigma_1 \sigma_2 + 6 \lambda_2 \sigma_2^2 + 2\lambda_2 \rho^2, \label{secvarpic1}\\
{\cal V}_{10} &\equiv & 2 \rho \bigl( (\lambda_3 - \lambda_4)\sigma_1 + 2\lambda_6\sigma_2\bigr), \no {\cal V}_{20} &\equiv & 2 \rho (\lambda_6 \sigma_1 + 8\lambda_2\sigma_2), \no {\cal V}_{00} &\equiv & 2(-\Delta_{22}+(\lambda_3 - \lambda_4) \sigma_1^2 + \lambda_6 \sigma_1\sigma_2 + 2\lambda_2 \sigma_2^2+ 6\lambda_2\rho^2). \label{secvarpic2}
\ea
The last element is equal to $8\lambda_2 \rho^2$ in the region where SPB takes place.
By taking one derivative w.r.t. $\mu$ of the gap equations and solving for
$\partial_\mu\sigma_j$ and $\partial_\mu\rho$ we find
\be
\partial_\mu\sigma_1 = - 4 {\cal N} \sigma_1\sqrt{\mu^2-\sigma_1^2}
\;\frac{{\cal V}_{22}{\cal V}_{00}- {\cal V}^2_{20}}{{\rm Det}{\cal V}},\qquad
\partial_\mu\sigma_2 = - 4 {\cal N} \sigma_1\sqrt{\mu^2-\sigma_1^2}
\;\frac{{\cal V}_{10}{\cal V}_{20}- {\cal V}_{12}{\cal V}_{00}}{{\rm Det}{\cal V}}, \label{eqmurho1}
\ee
\be
\partial_\mu\rho = - 4 {\cal N} \sigma_1\sqrt{\mu^2-\sigma_1^2}
\;\frac{{\cal V}_{12}{\cal V}_{20}- {\cal V}_{10}{\cal V}_{22}}{{\rm Det}{\cal V}}, \label{eqmurho}
\ee
At $\mu\to\infty$ the v.e.v have the following asymptotics:
$\sigma_1\sim 1/\mu^2$, $\sigma_2\sim {const}$, but these hold in a region where
our effective theory is not reliable anymore.

Let us compare the derivatives of the dynamic mass $\sigma_1$ across the phase transition point.
Their difference reads
\be
\partial_\mu\sigma_1\Big|_{\mu_{crit} +i0}- \partial_\mu\sigma_1\Big|_{\mu_{crit} -i0} =
- 4 {\cal N} \sigma_1\sqrt{\mu^2-\sigma_1^2}\frac{\big({\cal V}_{10}{\cal  V}_{22}-
{\cal V}_{20} {\cal V}_{12}\big)^2}{{\rm Det}{\cal V}\, {\rm Det}V^{(2)}_{\sigma}}\Big|_{\rho \rightarrow 0} < 0,
\ee provided that the determinants are positive (they determine the spectrum of meson masses squared).
Thus the derivative of the dynamic mass is discontinuous and the phase transition is of the second order.

We notice that convexity around this minimum implies that all diagonal elements are non-negative.
This gives positive masses for two scalar and four pseudoscalar mesons, whereas the doublet of
charged  of $\pi^\prime$ mesons remains massless. Quantitatively the mass spectrum can be obtained
only after kinetic terms are normalized. We just note that in the SPB phase the situation is rather
peculiar: pseudoscalar states mix with scalar ones. In particular, the diagonalization of kinetic
terms is different for neutral and charged pions because the vector isospin symmetry is broken:
$SU(2)_V \rightarrow U(1)$. This triggers a rather exotic mechanism of isospin breaking
via different decay constants. Even in the massless pion sector the isospin breaking
$SU(2)_V \rightarrow U(1)$ occurs: neutral pions become less stable with a larger decay
constant. We refer the reader to \cite{ae} for details.

SPB also induces mixing of both massless and heavy neutral pions with scalars. In fact in the
SPB phase parity is no longer a conserved quantity in strong interactions, so the distinction
between scalars and pseudoscalars is immaterial. This is why while the global broken
symmetry at the point of transition to the SPB is a vector one, the two Goldstone
bosons are apparently pseudoscalars, but as emphasized the distinction is purely semantic
once parity is broken.

\section{A particular example}

So far there is no sufficient experimental information to be able to fully determine
the value of the nine low energy constants appearing in $V_{\text{eff}}$. As we have seen
in section \ref{landscape} we can always, by means of a field redefinition, eliminate
$\lambda_5$ and diagonalize $\Delta_{ij}$. While using this freedom is quite convenient
to discuss the landscape of extrema of the effective potential, this transformation does
have some effects however in the coupling \eqref{yukawa} to quarks
$\bar q \sigma_1 q \rightarrow \bar q (\sigma_1 + \gamma \sigma_2) q$, which complicates
the discussion when introducing the chemical potential.

However, just to see that the emergence of the SPB phase is quite plausible, let us assume
that the low-energy model of QCD \eqref{effpot1} indeed has $\lambda_5=0$ and a diagonal
matrix $\Delta_{ij}$ whose eigenvalues need not be degenerate or, equivalently,
we take $\gamma = 0$ for the above vertex. Then following the procedure outlined in
section \ref{landscape} we can collect the set of inequalities ensuring the existence of
the two minima necessary for the emergence of the saturation point
\be
  \lambda_{1,2} > 0 ,\ \lambda_6 < 0,\ \Delta_{11}>0,\ \Delta_{22} > 0,\
(\lambda_3\pm\lambda_4) \Delta_{11}  > 2\lambda_1 \Delta_{22} ,\ (\lambda_3 +
\lambda_4)\Delta_{22} > 2\lambda_2 \Delta_{11} , \ x_1> \max(
-\frac{4\lambda_4}{\lambda_6}, -\frac38\frac{\lambda_6}{\lambda_2})
  \label{ineqq2}
\ee
If in addition we want to have SPB we have to require that  $\lambda_3, \lambda_4$ are positive
(see \cite{ae}). The sign of $\sigma_1$ is fixed by current algebra considerations
thus removing the degeneracy of solutions.

Both sets of solutions, namely these with $\sigma_2 =0$ or these with $\sigma_2 \not=0$
may provide the lowest minimum at $\mu = 0$. This turns out to be controlled by the sign
of the combination $ 8 \lambda_2 \lambda_4 -\lambda_6^2$ . The latter follows from
the combined analysis of the vacuum solutions and the solutions at the SPB phase transition
(see Sec. \ref{SPB}). It is clear that the conditions leave a lot of room for the simultaneous
occurrence of both phenomena. In \cite{ae} a rather rough phenomenological fit was done leading
e.g. to $\lambda_1 \sim 0.15, \lambda_3 \sim 4, \Delta_{11} \sim 0.03$ GeV$^{-2}$,
$\Delta_{22} \sim 0.1$ GeV$^{-2}$ which would trigger SPB at about three normal nuclear densities.
Nuclear matter is formed at $\mu_* \simeq 303$ MeV stabilized by a $\omega$ meson condensate
with $G_\omega \sim (10 \div 15)$ GeV$^{-2}$ in satisfactory agreement with other
model estimations \cite{vectorm}.

\section{Conclusions}
Let us summarize here our main findings. Parity violation seems to be quite a realistic possibility in nuclear matter at moderate densities. We have arrived at this conclusion by using an effective Lagrangian for low-energy QCD that retains the two lowest lying states in the scalar and pseudoscalar sectors. We include a chemical potential for the quarks that corresponds to a finite density of baryons, implement the bound state of normal nuclear matter and investigate the pattern of symmetry violation in its presence. We have found the necessary and sufficient conditions for a phase where parity is spontaneously broken to exist. In general this phase is bound and it extends across a range of chemical potentials that correspond to nuclear densities where more exotic phenomena such as CFL or CS may occur.

Salient characteristics of this phase would be the spontaneous violation of the vector isospin symmetry $SU(2)_V$ down to $U(1)$ and the generation two additional massless charged pseudoscalar mesons. We also find a strong mixing between scalar and pseudoscalar states that translate spontaneous parity violation into meson decays. The mass eigenstates will decay both in odd and even number of pions simultaneously. Isospin violation can also be visible in decay constants.

We think that our conclusions are drawn in a region of parameters where effective Lagrangian techniques are applicable and, while obviously we cannot claim high accuracy in our predictions,  we are confident that the existence of this novel phase is not an spurious consequence of our approach but a rather robust prediction. It would surely be interesting to investigate how this new phenomenon could possibly influence the equation of state of neutron stars .
%(the density of such objects seems to be about right for it).

One could hope that lattice methods \cite{philip} may shed some light on this issue and confirm or falsify the existence of this interesting phase in dense nuclear matter. Work along this lines is proceeding \cite{aep}.

\section*{Acknowledgments} This work was supported by research
grants FPA2007-66665, 2005SGR00564, 2009SGR and RFBR 09-02-00073-a.
It is also supported by the Consolider-Ingenio 2010 Program CPAN (CSD2007- 00042). We acknowledge the partial support of the EU RTN networks FLAVIANET and ENRAGE and the Program RNP2009-1575. D.E. wishes to thank the PH division of CERN where this work was partly done for the hospitality extended to him.

%\begin{figure}[h]
%\epsfysize=3cm
%\epsfbox{andrianov.eps}
%\caption{Evolution of the scalar and pseudoscalar v.e.v. %$\sigma_1(\mu)$
%and $\rho(\mu)$, respectively, as a function of the baryon chemical %potential.
%\label{fig1}}
%\end{figure}


\begin{thebibliography}{99}

\bibitem{migdal} A.B. Migdal, Zh. Eksp. Teor. Fiz. 61 (1971) 2210 [Sov. Phys.
JETP 36 (1973) 1052];\, R.F. Sawyer, Phys. Rev. Lett. 29 (1972) 382;\, D.J. Scalapino, Phys. Rev. Lett. 29 (1972) 386;\, G. Baym, Phys. Rev. Lett. 30 (1973) 1340;\ A.B. Migdal, O.A. Markin and I.N. Mishustin, Sov. Phys. JETP, 39 (1974) 212.

\bibitem{picon} A.B. Migdal, Rev. Mod. Phys. 50 (1978) 107;\, D. Bailin and A.
Love, Phys. Rep. 107, 325 (1984);\, C.-H. Lee, Phys. Rep. 275 (1996) 197;\, M. Prakash, I. Bombaci, M. Prakash, P. J. Ellis, J. M. Lattimer and R. Knorren, Phys. Rep. 280 (1997) 1.

\bibitem{kharzeev} D. Kharzeev, R.D. Pisarski and M.H.G. Tytgat,
Phys. Rev. Lett. 81 (1998) 512;\, D. Kharzeev and R.D. Pisarski, Phys. Rev. D 61 (2000) 111901(R);\, D. Kharzeev, Phys. Lett. B 633 (2006) 260;\, D. Kharzeev and A. Zhitnitsky, Nucl. Phys.  A 797, 67 (2007) 67;\, [hep-ph];\, D.E. Kharzeev, L.D. McLerran and H.J. Warringa,
  Nucl. Phys.  A 803 (2008) 227;\,  H.J. Warringa,
  J. Phys. G 35, 104012 (2008) .

\bibitem{scon} M. Alford, K. Rajagopal and F. Wilczek, Phys. Lett. B422, 247
(1998);\, R. Rapp, T. Schaefer, E. V. Shuryak and M. Velkovsky, Phys. Rev. Lett. 81  (1998) 53 .

\bibitem{witten} D. Weingarten, Phys. Rev. Lett. 51, 1830 (1983);\, C. Vafa and
E. Witten, Phys. Rev. Lett. 53 (1984) 535;\,S. Nussinov, Phys. Rev. Lett. 52, 966 (1984);\, D. Espriu, M. Gross and J.F. Wheater, Phys. Lett. B 146, 67 (1984) .

\bibitem{ae} A. Andrianov and D. Espriu, Phys. Lett. B 663, 450 (2008).

\bibitem{buballa1} M. Buballa, Nucl. Phys. A 611 (1996) 393.

\bibitem{aae} A. Andrianov, V. Andrianov and D. Espriu, in preparation.

\bibitem{bary} G.E. Brown, M. Rho, Phys. Rep. 363 (2002) 85;\, D. Toublan and J.
B. Kogut, Phys. Lett. B 564, 212 (2003);\,  M. Frank, M. Buballa and M. Oertel, Phys. Lett. B 562, 221 (2003).

\bibitem{takah} K. Takahashi, Phys. Rev. C 66 (2002) 025202

\bibitem{sigqu} D. Bailin, J. Cleymans and M.D. Scadron, Phys. Rev. D 31, 164
(1985);\,   O. Scavenius, \'A. M\'ocsy, I.N. Mishustin and D.H. Rischke, Phys. Rev. C 64, 045202 (2001);\, P.A.M. Guichon and A.W. Thomas, Phys. Rev. Lett. 93 (2004) 132502;\,  S. Lawley, W. Bentz and A.W. Thomas, J. Phys. G 32 (2006) 667 .

\bibitem{grei} S. Pal, M. Hanauske, I. Zakout, H. Stoecker and W. Greiner, Phys.
Rev. C, 60, 015802 (1999).

\bibitem{NJL} V. Bernard, Ulf-G.~Meissner and I. Zahed, Phys. Rev D36 (1987)
819;\, M. Asakawa and K. Yazaki, Nucl. Phys. A 504 (1989) 668;\,  T. Hatsuda and T. Kunihiro, Phys. Rep., 247, 221 (1994);\, A. Delfino, J. Dey, M. Dey, M. Malheiro, Phys. Lett. B363, (1995) 17;\, M. Buballa, Phys. Rept. 407 (2005) 205.

\bibitem{ebert} A. Barducci, R. Casalbuoni, G. Pettini, and L. Ravagli, Phys.
Rev. D 69 (2004) 096004;\, D. Ebert and K.G. Klimenko, J.Phys. G32 (2006) 599;\, Eur.Phys.J. C46 (2006) 771.

\bibitem{philip} O. Philipsen,   Eur. Phys. J. ST 152 (2007) 29; \, M.P. Lombardo,
PoS CPOD2006 (2006) 003 ;\, J. Phys. G 35, 104019 (2008);\, M.A. Stephanov, PoS LAT2006 (2006) 024.

\bibitem{heavyion} P. Braun-Munzinger, K. Redlich and J. Stachel, arXiv: nucl-th/0304013;\,
H. Stocker, Conf. Proc. C0806233, moycgm01 (2008);\, G.V. Trubnikov {\it et al.}, Conf. Proc. C0806233, WEPP029 .

\bibitem{gellmann} M. Gell-Mann and M. Levy, Nuovo Cim. 16 (1960) 705.

\bibitem{wal} B. D. Serot and J. D. Walecka, Adv. Nucl. Phys. 16 (1986), 1; Int. J. Mod. Phys. E 16
(1997), 15.

\bibitem{vectorm} \'E. Massot and G. Chanfray, Phys. Rev.  C 78, 015204 (2008);\ R. Huguet, J.C. Caillon and J. Labarsouque, Nucl. Phys.  A {\bf 809} (2008) 189 and refs. therein.



\bibitem{aep} A. Andrianov, D. Espriu and A. Papa, in preparation.


\end{thebibliography}
\end{document}